\documentstyle[12pt]{article}
\begin{document}
\begin{center}
\vfill
\large\bf{A Functional Approach to the}\\
\large\bf{Heat Kernel in Curved Space}
\end{center}
\vfill
\begin{center}
L. Martin\\
D.G.C. McKeon\\
Department of Applied Mathematics\\
University of Western Ontario\\
London\\
CANADA\\
N6A 5B7
\end{center}
\vfill
email: LCM@APMATHS.UWO.CA\\
TMLEAFS@APMATHS.UWO.CA\\
Tel: (519)679-2111, ext. 8789\\
Fax: (519)661-3523\\
\hfill                     PACS No.: 11.20Dj
\eject

\section{Abstract}
The heat kernel $M_{xy} = <x\mid \exp \left[ \frac{1}{\sqrt{g}} 
\partial_\mu g^{\mu\nu}
\sqrt{g}
\partial_\nu \right]t \mid y>$ is of central importance when studying 
the propagation of a scalar
particle in curved space.  It is quite convenient to analyze this 
quantity in terms of classical
variables by use of the quantum mechanical path integral; regrettably 
it is not entirely clear how
this path integral can be mathematically well defined in curved 
space.  An alternate approach to
studying the heat kernel in terms of classical variables was 
introduced by Onofri.  This technique
is shown to be applicable to problems in curved space; an unambiguous 
expression for $M_{xy}$
is obtained which involves functional derivatives of a classical 
quantity.  We illustrate how this
can be used by computing $M_{xx}$ to lowest order in the curvature 
scalar $R$.
\section{Introduction}
The heat kernel $<x \mid e^{-Ht} \mid y>$ for an elliptic operator 
$H$ is a fundamental quantity
in quantum field theory [1,2].  Its expansion in powers of $t$ can be 
used to analyze the
divergence structure of a theory [2,3]; one can also expand in powers 
of the background field on
which $H$ depends to obtain Green's functions [1,4].

In both cases, the quantum mechanical path integral is a useful means 
of representing the heat
kernel, for then only classical variables need to be manipulated in 
performing calculations [5,6,7].
At a practical level, when computing Green's functions using this 
technique, loop-momentum
integrals are avoided and algebraicly complicated vertices in gauge 
theories are not encountered. 
The application of the quantum mechanical path integral to problems 
in curved space [8,9] is
unfortunately hampered by the difficulties encountered in defining 
unambiguously the quantum
mechanical path integral in curved space [10-13].

Fortunately, an alternate way of representing the heat kernel in 
terms of classical variables can
be developed using an approach initiated by Onofri [14].  This 
technique has been shown to be
a viable way of computing Green's functions in flat space by using it 
to evaluate the one-loop
vacuum polarization in scalar electrodynamics [15].  It retains all 
the advantages of the quantum
mechanical path integral.  In this paper, it is shown that one can 
use Onofri's method to handle
the heat kernel in $D$ dimensional curved space that arises when one 
considers the propagation
of scalars in which case
$$H = \frac{1}{2} \frac{1}{\sqrt{g}} p_\mu (g^{\mu\nu} \sqrt{g}) 
p_\nu \; .\;\;\;\; (p = -
i\partial)\eqno(1)$$
We use a normal coordinate expansion [16] of the metric $g_{\mu\nu}$, 
following the approach
initially used in the non-linear sigma model [17].  The general 
formalism is developed, and the
diagonal elements of the heat kernel is computed to leading order in 
the Riemann scalar $R$.

\section{The Heat Kernel}
We begin with the normal coordinate expansion of the metric tensor 
$g_{\mu\nu}$ about a point
$x_0$ where the metric is flat [16]:
$$g_{\mu\nu} (x_0 + \pi(q))  =  \eta_{\mu\nu} + \frac{1}{3} 
R_{\mu\alpha\beta\nu} (x_0)
q^\alpha q^\beta  + \frac{1}{6} R_{\mu\alpha\beta\nu;\gamma} (x_0) 
q^\alpha q^\beta
q^\gamma\nonumber$$
$$ +  \left( \frac{1}{20} R_{\mu\alpha\beta\nu;\gamma\delta} (x_0) + 
\frac{2}{45} R_{\mu \alpha
\beta\sigma} (x_0) R^\sigma_{\hspace{.2cm}\gamma\delta\nu} (x_0) 
\right) q^\alpha q^\beta
q^\gamma
q^\delta
 + \ldots \eqno(2)$$
so that $H$ defined in (1) becomes, to at most fourth order in 
$y^\alpha$,
$$H = \frac{1}{2} \left[ 1 + \frac{1}{6} R_{\alpha\beta} q^\alpha 
q^\beta + \frac{1}{12}
R_{\alpha \beta; \gamma} q^\alpha q^\beta q^\gamma + \left( 
\frac{1}{72} R_{\alpha\beta}
R_{\gamma\delta} + \frac{1}{40} R_{\alpha\beta; \gamma\delta} 
\right.\right.\nonumber$$
$$\left.\left. + \frac{1}{180} 
R^\lambda_{\hspace{.2cm}\alpha\beta\sigma}
R_{\lambda\gamma\delta}^{\hspace{.4cm}\sigma}
\right) q^\alpha q^\beta q^\gamma q^\delta \right] p_\mu \left[ 
\eta^{\mu\nu} + \left(- \frac{1}{6}
R_{\alpha\beta} \eta^{\mu\nu} \right.\right. \nonumber$$
$$\left. + \frac{1}{3}
R_{ {\hspace{.3cm}\alpha}{\hspace{.5cm}\beta} 
}^{\mu{\hspace{.4cm}\nu}} \right) q^\alpha q^\beta  + \left( -
\frac{1}{12}
\eta^{\mu\nu} R_{\alpha\beta ; \gamma} + \frac{1}{6}
R^{\mu\hspace{.3cm}\nu}_{{\hspace{.2cm}\alpha}{\hspace{.2cm}\beta;\gamma}}
\right)q^\alpha q^\beta q^\gamma\nonumber$$
$$+ \left( \eta^{\mu\nu} \left( \frac{1}{72} R_{\alpha\beta} 
R_{\gamma\delta} -
\frac{1}{20} R_{\alpha\beta ; \gamma\delta} - \frac{1}{90}
R^\lambda_{\hspace{.2cm}\alpha\beta\kappa}R_{\lambda\gamma\delta}^
{\hspace{.5cm}\kappa}\right.\right) \nonumber$$
$$ \left. \left. \frac{1}{20}
R_{{\hspace{.2cm}\alpha}{\hspace{.4cm}\beta;
\gamma\delta}}^{\mu\hspace{.3cm}\nu} 
+ \frac{1}{15}
R_{\alpha{\hspace{.4cm}\beta}}^{{\hspace{.3cm}\mu}{\hspace{.6cm}\kappa}}
R_{\kappa\gamma{\hspace{.4cm}\delta}}^{\hspace{.4cm}\nu} \right) 
q^\alpha
q^\beta q^\gamma
q^\delta \right] p_\nu\;\; . \eqno(3)$$
(We define $R \equiv R^\alpha_{\hspace{.2cm}\alpha}$ where 
$R_{\alpha\beta} \equiv
R^\mu_{\hspace{.2cm}{\alpha\mu\beta}}$.)  If we now express $H$ in 
the form
$$H = \frac{1}{2} \left[ p^2 + M(q) p^2 + p_\mu N^{\mu\nu} (q) p_\nu 
+ M(q) p_\mu
N^{\mu\nu} (q) p_\nu \right]\;\; ,\eqno(4)$$
then it is possible to employ the techniques of [14], as used in 
[15].  We first use the relation
$$e^{A+B} = e^A T \exp \int_0^1 d\tau \left[e^{-A\tau} B e^{A\tau} 
\right]\eqno(5)$$
$$(T - {\rm{path\;\;ordering}})\nonumber$$
to write the matrix element $M_{xy}$ as
$$<x \mid e^{-Ht} \mid y>\; = \;<x \mid e^{-\frac{1}{2} p^2 t} T \exp 
- \frac{t}{2} \int_0^1
d\tau
\left[ e^{\frac{1}{2} p^2 t \tau} \left(M p^2 + p_\mu N^{\mu\nu} 
p_\nu \right.\right.\nonumber$$
$$\left.\left. +\; M p_\mu N^{\mu\nu} p_\nu \right) e^{-\frac{1}{2} 
p^2 t \tau} \right] \mid y
>\;\;
. \eqno(6)$$
Since we have the identity
$$e^A f(B) e^{-A} = f(e^{[A}, B)\nonumber$$
$$= f(B + \frac{1}{1!} [A,B] + \frac{1}{2!}[A,[A,B]] + \ldots )\;\; , 
\eqno(7)$$
we can rewrite $M_{xy}$ in (6) as
$$M_{xy} = <x \mid e^{-\frac{1}{2} p^2 t} T \exp \frac{-t}{2} 
\int_0^1 d\tau \left[ M(q - it \tau
p) p^2 + p_\mu N^{\mu\nu} (q - it \tau p)p_\nu \right. \nonumber$$
$$\left. + M(q - it \tau p) p_\mu N^{\mu\nu} (q - it \tau p) p_\nu 
\right] \mid y>\eqno(8)$$
since
$$[q,p] = i\;\; .\eqno(9)$$
Insertion of appropriate complete sets of states into (8) now gives us
$$M_{xy} = \frac{1}{(2\pi t)^{D/2}} \int dx^\prime < x^\prime \mid 
e^{-(x-q)^2/2t} T \exp -
\frac{t}{2} \int_0^1 d\tau \left[M(q - i t \tau p) p^2\right. 
\nonumber$$
$$\left. + p_\mu N^{\mu\nu} (q - i t \tau p) p_\nu + M(q - i t \tau 
p)p_\mu N^{\mu\nu} (q-i t
\tau p)p_\nu\right]\nonumber$$
$$e^{(x-q)^2/2t} \mid y> e^{-(x-y)^2/2t} \;\; .\eqno(10)$$
Again employing (7), we rewrite (10) in the form
$$M_{xy} = \frac{e^{-(x-y)^2/2t}}{(2\pi t)^{D/2}} \int dx^\prime < 
x^\prime \mid T \exp -
\frac{t}{2} \int_0^1 d\tau \left[M(q(1-\tau) - i t \tau p + \tau 
x)\left(p + \frac{i(x-q)}{t}\right)^2
\right. \nonumber$$
$$\left. + \ldots \right] \mid y>\eqno(11)$$
$$\;\;\;= \frac{e^{-(x-y)^2/2t}}{(2\pi t)^{D/2}}<p = 0 \mid 
e^{ip\cdot y} T \exp - \frac{t}{2}
\int_0^1
d\tau \left[M(q(1-\tau) - i t \tau p + \tau x)\left(p + \frac{i(x-
q)}{t}\right)^2 \right. \nonumber$$
$$\left. + \ldots \right] e^{-i p\cdot y} \mid q = 0 > \;\; 
.\eqno(12)$$
A last application of (7) converts (12) to
$$M_{xy} = \frac{e^{-(x-y)^2/2t}}{(2\pi t)^{D/2}}<p = 0 \mid T \exp - 
\frac{t}{2} \int_0^1
d\tau
\left[M (y(\tau) + q(1 - \tau) - i t \tau p) \right.\nonumber$$
$$ \left( \Delta + p - \frac{iq}{t} \right)^2 + \left( \Delta + p - 
\frac{iq}{t} \right)_\mu
N^{\mu\nu} \left(y(\tau) + q(1-\tau) - i t \tau p \right)\nonumber$$
$$ \left( \Delta + p - \frac{iq}{t} \right)_\nu + M \left(y(\tau) + 
q(1-\tau) - i t \tau p \right) \left(
\Delta + p - \frac{iq}{t} \right)_\mu \nonumber$$
$$\left. N^{\mu\nu} \left(y(\tau) + q(1-\tau) - i t \tau p\right) 
\left(\Delta + p - \frac{iq}{t}
\right)_\nu \right] \mid q = 0> \eqno(13)$$
where $y(\tau) = y + (x-y) \tau$ and $\Delta = \frac{i}{t} (x-y)$.  
An expansion in (13) of the
functions $M$ and $N^{\mu\nu}$ about $y(\tau)$ followed by an 
expansion of the path-ordered
exponential along the lines of [14] leads to
$$M_{xy} = \frac{ e^{-(x-y)^2/2t}}{(2\pi t)^{D/2}} \sum_{N=0}^\infty 
\left( - \frac{t}{2}
\right)^N
\int_0^1 d\tau_1 \int_0^{\tau_1} d\tau_2 \ldots \int_0^{\tau_{N-1}} 
d\tau_N\nonumber$$
$$\left\lbrace \left[ M\left(y(\tau_1) + 
\frac{\delta}{\delta\alpha(\tau_1)} \right) \left( \Delta +
\frac{\delta}{\delta\xi(\tau_1)} \right) \cdot \left(\Delta + 
\frac{\delta}{\delta \zeta
(\tau_1)}\right)\right.\right.\nonumber $$
$$ + \left( \Delta + \frac{\delta}{\delta\xi(\tau_1)} \right)_\mu 
N^{\mu\nu} \left( y(\tau_1) +
\frac{\delta}{\delta \beta(\tau_1)}\right) \left( \Delta + 
\frac{\delta}{\delta\zeta(\tau_1)}
\right)_\nu\nonumber$$
$$\left. + M\left(y(\tau_1) + \frac{\delta}{\delta\alpha(\tau_1)} 
\right) \left( \Delta +
\frac{\delta}{\delta \xi (\tau_1)} \right)_\mu N^{\mu\nu} \left( 
y(\tau_1) +
\frac{\delta}{\delta\beta(\tau_1)} \right)\left( \Delta + 
\frac{\delta}{\delta\zeta(\tau_1)} \right)_\nu
\right]\nonumber$$
$$\ldots \left[ M \left(y(\tau_N) +  
\frac{\delta}{\delta\alpha(\tau_N)} \right) \left( \Delta +
\frac{\delta}{\delta\xi(\tau_N)} \right) \cdot \left( \Delta + 
\frac{\delta}{\delta\zeta(\tau_N)}
\right)\right.\nonumber$$
$$ + \left( \Delta + \frac{\delta}{\delta\xi(\tau_N)} \right)_\mu 
N^{\mu\nu} \left(y (\tau_N) +
\frac{\delta}{\delta\beta(\tau_N)} \right) \left( \Delta + 
\frac{\delta}{\delta\zeta(\tau_N)}
\right)_\nu\nonumber$$
$$\left. + M \left( y(\tau_N) +  \frac{\delta}{\delta\alpha(\tau_N)} 
\right) \left( \Delta +
\frac{\delta}{\delta\xi(\tau_N)} \right)_\mu N^{\mu\nu} 
\left(y(\tau_N) + \frac{\delta}{\delta 
\beta(\tau_N)} \right)\left( \Delta + 
\frac{\delta}{\delta\zeta(\tau_N)} \right)_\nu
\right]\nonumber$$
$$< p = 0 \mid e^{\alpha(\tau_1)\cdot (I\!\!P_1^a + Q_1^a)} 
e^{\xi(\tau_1)\cdot (I\!\!P_1^b +
Q_1^b)} e^{\beta(\tau_1)\cdot (I\!\!P_1^a + Q_1^a)} 
e^{\zeta(\tau_1)\cdot (I\!\!P_1^b +
Q_1^b)}\nonumber$$
$$\ldots e^{\alpha(\tau_N) (I\!\!P_N^a + Q_N^a)} e^{\xi(\tau_N)\cdot 
(I\!\!P_N^b + Q_N^b)}
e^{\beta(\tau_N)\cdot (I\!\!P_N^a + Q_N^a)} e^{\zeta(\tau_N)\cdot 
(I\!\!P_N^b + Q_N^b)} \mid
q = 0 >\eqno(14)$$
when $\alpha = \beta = \xi = \zeta = 0$.  We have defined
$$Q_i^a = q(1 - \tau_i)\eqno(15)$$
$$I\!\!P_i^a = -i t \tau_i p\eqno(16)$$
$$Q_i^b = - i q/t\eqno(17)$$
and
$$I\!\!P_i^b = p\;\; .\eqno(18)$$
If now $[Q_i^c, I\!\!P_j^d] = {\displaystyle{C_{(i,j)}^{(c,d)}}}$ with
${\displaystyle{C^{(s)\,\,\,(c,d)}_{\;\;\;\;\;\;(i,j)}}} = 
{\displaystyle{C_{(i,j)}^{(c,d)}}}\;\;(i < j)$
and
${\displaystyle{C^{(s)\,\,\,(c,d)}_{\;\;\;\;\;\;(i,j)}}} = 
{\displaystyle{C_{(j,i)}^{(d,c)}}}$, then we
can use the result from (14);
$$< p = 0 \mid e^{\lambda_1(I\!\!P_1 + Q_1)} \ldots 
e^{\lambda_N(I\!\!P_N + Q_N)} \mid q
= 0>\nonumber$$
$$\hspace{.8cm} = \exp \frac{1}{2} \sum_{i;j} \left(C^{(s)}_{ij} 
\lambda_i \lambda_j
\right)\;\;.\eqno(19)$$
This leaves us with
$$M_{xy} = \frac{e^{-(x-y)^2/2t}}{(2\pi t)^{D/t}} \exp \frac{1}{2} 
\int_0^1 dt_1  \int_0^1 dt_2
\left[ t t_< (1 - t_>) \left(\frac{\delta}{\delta\alpha(t_1)} \cdot 
\frac{\delta}{\delta\alpha(t_2)}   
\right. \right.\nonumber$$
$$ \left. + 2 \frac{\delta}{\delta\alpha(t_1)}\;\; 
\frac{\delta}{\delta\beta(t_2)} +
\frac{\delta}{\delta\beta(t_1)} \cdot 
\frac{\delta}{\delta\beta(t_2)}\right) + \frac{1}{t}\left(
\frac{\delta}{\delta\xi(t_1)}\cdot \frac{\delta}{\delta\xi(t_2)}   
\right. \nonumber$$
$$ \left. + 2 \frac{\delta}{\delta\xi(t_1)}\cdot 
\frac{\delta}{\delta\zeta(t_2)} +
\frac{\delta}{\delta\zeta(t_1)} \cdot 
\frac{\delta}{\delta\zeta(t_2)}\right) + 2i \left(\theta_1(t_1 -
t_2) - t_1\right) \nonumber$$
$$ \left( \frac{\delta}{\delta\alpha(t_1)}\cdot 
\frac{\delta}{\delta\xi(t_2)} +
\frac{\delta}{\delta\alpha(t_1)} \cdot 
\frac{\delta}{\delta\zeta(t_2)} +
\frac{\delta}{\delta\beta(t_1)}\cdot 
\frac{\delta}{\delta\zeta(t_2)}\right) \nonumber$$
$$\left. + 2i \left( \theta_0 (t_1 - t_2) - t_1\right) 
\left(\frac{\delta}{\delta\beta(t_1)} \cdot
\frac{\delta}{\delta\xi(t_2)} \right) \right]\eqno(20)$$
$$\exp - \frac{t}{2} \int_0^1 d\tau \left[ M(y (\tau) + \alpha(\tau)) 
(\Delta + \xi(\tau)) \cdot
(\Delta + \zeta(\tau))\right.\nonumber$$
$$+ (\Delta + \xi(\tau))_\mu N^{\mu\nu} (y(\tau) + \beta(\tau)) 
(\Delta +
\zeta(\tau))_\nu\nonumber$$
$$\left. + M(y(\tau) + \alpha(\tau)) (\Delta + \xi(\tau))_\mu 
N^{\mu\nu} (y(\tau)
+\beta(\tau))(\Delta + \zeta(\tau))_\nu \right]\; , \nonumber$$
where $t_< = \min (t_1, t_2), t_> = \max (t_1, t_2)$ and 
$\theta_a(\tau) = 1 (\tau > 0), = 0 (\tau
< 0), = a (\tau = 0)$.  In (20), we have an unambiguous closed form 
expression for the heat
kernel in terms of classical variables, a feature it shares with the
path integral representation of the heat kernel.

For purposes of illustration, let us compute the contribution to 
$M_{xx}$ that is linear in the
Riemann scalar $R$.  Those terms in (20) needed for this are
$$M_{xx}^{(R)} = \frac{1}{(2\pi t)^{D/2}} \left( \frac{1}{2!} (2) 
(\frac{1}{2})^2 \right)
\int_0^1 dt_1 dt_2 \int_0^1 dt_1^\prime dt_2^\prime\nonumber$$
$$\left\lbrace \left[t t_<(1-t_>) \left( 
\frac{\delta}{\delta\alpha(t_1)}\cdot
\frac{\delta}{\delta\alpha(t_2)} + 2 
\frac{\delta}{\delta\alpha(t_1)}\cdot
\frac{\delta}{\delta\beta(t_2)} + 
\frac{\delta}{\delta\beta(t_1)}\cdot \frac{\delta}{\delta\beta(t_2)} 
\right)\right]  \right.\nonumber$$
$$\left[\frac{1}{t} \left(\frac{\delta}{\delta\xi(t_1^\prime)}\cdot
\frac{\delta}{\delta\xi(t_2^\prime)} + 2 
\frac{\delta}{\delta\xi(t_1^\prime)} \cdot
\frac{\delta}{\delta\zeta(t_2^\prime)} + 
\frac{\delta}{\delta\zeta(t_1^\prime)}\cdot
\frac{\delta}{\delta\zeta(t_2^\prime)}\right) \right] \nonumber$$
$$+ (2i)^2 \left[(\theta_1(t_1 - t_2) - t_1)
\left(\frac{\delta}{\delta\alpha(t_1)}\cdot\frac{\delta}{\delta\xi(t_2)}
\frac{\delta}{\delta\alpha(t_1)} \cdot 
\frac{\delta}{\delta\zeta(t_2)} +
\frac{\delta}{\delta\beta(t_1)}\cdot 
\frac{\delta}{\delta\zeta(t_2)}\right) \right.\nonumber$$
$$\left. + (\theta_0 (t_1 - t_2) - t_1) \left( 
\frac{\delta}{\delta\beta(t_1)}\cdot
\frac{\delta}{\delta\xi(t_2)}\right)\right]\nonumber$$
$$\left[(\theta_1(t_1^\prime - t_2^\prime) - t_1^\prime)
\left(\frac{\delta}{\delta\alpha(t_1^\prime)}\cdot\frac{\delta}{\delta}
\frac{\delta}{\delta\alpha(t_1^\prime)} \cdot 
\frac{\delta}{\delta\zeta(t_2^\prime)} +
\frac{\delta}{\delta\beta(t_1^\prime)}\cdot 
\frac{\delta}{\delta\zeta(t_2^\prime)}\right)
\right.\nonumber$$
$$\left. \left. + (\theta_0(t_1^\prime - t_2^\prime) - t_1^\prime)
\left(\frac{\delta}{\delta\beta(t_1^\prime)}\cdot 
\frac{\delta}{\delta\xi(t_2^\prime)}\right)
\right]\right\}\eqno(21)$$
$$\left\{ (- \frac{t}{2}) \int_0^1 d\tau \left[ \frac{1}{6} 
R_{\alpha\beta} \alpha^\alpha
(\tau)\alpha^\beta(\tau) \xi(\tau)\cdot \zeta(\tau)\right. 
\right.\nonumber$$
$$\left.\left.\left. + (- \frac{1}{6} R_{\alpha\beta} \eta^{\mu\nu} + 
\frac{1}{3}
R_{{\hspace{.3cm}\alpha}{\hspace{.4cm}\beta}}^{\mu{\hspace{.4cm}\nu}}
 \right) \beta^\alpha (\tau) \beta^\beta
(\tau)\xi_\mu (\tau)\zeta_\nu(\tau)\right]
\right\}\;\; . \nonumber$$
It is a straightforward exercise to compute the functional 
derivatives in (21) and to then evaluate
all remaining integrals to obtain 
$$M_{xx}^{(R)} = \frac{1}{(2 \pi t)^{D/2}} \left( \frac{Rt}{12} 
\right) \eqno(22)$$
in agreement with [2,9].  Further contributions can be similarly 
evaluated.

\section{Discussion}
The heat kernel associated with the propagation of a scalar particle 
in curved space can be
expressed in terms of classical variables in several ways.  One 
involves writing the heat kernel
in terms of the quantum mechanical path integral; this unfortunately 
is not a straightforward
exercise and a number [10-13] of different effective actions for the 
path integral have been
proposed.  (Those effective actions which are derived by a non-
covariant time-slicing are in fact
not covariant themselves.)  An alternate approach to expressing the 
heat kernel in curved space
in terms of classical variables has been developed in the preceding 
section.  It is
unambiguous and
easily used in practical calculations as it only involves evaluating 
functional derivatives of an
effective action and computing some elementary
integrals.  We shall endeavour to illustrate how this approach can be 
used to facilitate higher loop
corrections in quantum gravity.

\section{Acknowledgements}

We would like to thank F. Dilkes for useful discussion.  NSERC 
provided financial support.  R.
and D. MacKenzie made a useful suggestion and M. Harris provided
motivation for this work.

\end{document}